\documentclass[aps,prl,twocolumn,10pt]{revtex4}

\usepackage{xcolor, graphics, graphicx, epsfig, latexsym, amsmath, lmodern, braket}
\usepackage{dcolumn}   % needed for some tables

\DeclareMathOperator{\Tr}{Tr}

\begin{document}

\title{Precision and reproducibility of macroscopic developmental patterns}

\author{Laurent Abouchar*, Mariela D. Petkova*, Cynthia R. Steinhardt, and Thomas Gregor}
\affiliation{Joseph Henry Laboratories of Physics, and Lewis Sigler Institute for Integrative Genomics, Princeton University, New Jersey 08544, USA}

\begin{abstract}
Developmental processes in multicellular organisms occur far from equilibrium, yet produce complex patterns with astonishing reproducibility. We measure the precision and reproducibility of bilaterally symmetric fly wings across the natural range of genetic and environmental conditions and find that wing patterns are specified with identical spatial precision and are reproducible to within a single cell width. The early fly embryo operates at a similar degree of reproducibility, suggesting that the overall spatial precision of morphogenesis in {\it Drosophila} performs at the single cell level, arguably the physical limit of what a biological system can achieve. 
\end{abstract}

\maketitle

%
%
%
%%%% INTRODUCTION
%
%

Physically, animate and inanimate systems are distinctly separable by one quantifiable aspect: their ability to generate reproducible complex patterns. Contrary to intuition, there are no two identical snowflakes while the bodies of insects are covered with patterns that are indistinguishable across individuals~\cite{Wigglesworth:1940, Maynard-Smith:1960, Lawrence:1973}. As illustrated by snowflakes, pattern formation occurs under non-equilibrium conditions: it is history-dependent and sensitive to initial conditions, rendering it difficult to predict final outcomes~\cite{Cross:1993}. As a general rule, pattern formation in biological processes such as morphogenesis proceeds sequentially, interpolating coarser patterns of previous processes to achieve finer differentiation~\cite{Thompson:1917, Cross:2009}. Despite the tendency for amplifying variability from one patterning layer to the next in addition to differences in the genetic and environmental conditions, complex final structures are formed with high reproducibility.  Morphogenesis has solved the reproducibility problem in non-equilibrium processes, but its strategy remains unclear.

The bilaterally symmetric wings of insects, such as the fruit fly, provide an ideal opportunity to identify quantitative rules of biological pattern formation. 
%Wing morphogenesis begins with as few as 6 cells in a 3h old embryo that undergo 12-13 divisions within 6 days to generate a disc in the fly larva with $\sim\!50,000$ cells. In the disc, the position of wing features are specified by spatially distributed molecular signals~\cite{Celis:2003}. 
%In particular, each of the two fly wings is generated from a disc in the fly larva where the position of wing features are specified by spatially distributed molecular signals~\cite{Celis:2003}.  Subsequently, the cells no longer divide and the interplay of these instructive molecular signals, individual cell connectivity properties~\cite{Classen:2005} and global mechanical deformations~\cite{Aigouy:2010} transform the disc into a highly stereotyped wing blade with a pattern of five longitudinal and two transverse veins (Fig.~1A). 
Fly wings have highly stereotyped blade structure with a pattern of five longitudinal and two transverse veins (Fig.~1A). Each wing is generated from a disc with $\sim$50,000 cells in the fly larva (Table I), where the positions of wing features are specified by spatially distributed molecular signals~\cite{Celis:2003}. The fly provides us with two independent handles for quantifying the patterning process: first, because the formation of the left and right wing is essentially a reiteration of the same process under identical conditions~\cite{VanValen:1962, Debat:2003}, the two wings are an internal control for the {\it precision} of the patterning process, i.e. a measure for bilateral symmetry. Second, in randomly chosen left and right wings from a population of flies, wing formation is enacted under different conditions for each individual, allowing for a measure of pattern {\it reproducibility}. Thus, we can carry out a performance analysis of wing morphogenesis and probe the limits of precision and reproducibility to assess developmental patterning fidelity.

\begin{figure}[t!]
\centering
\includegraphics[width=\columnwidth]{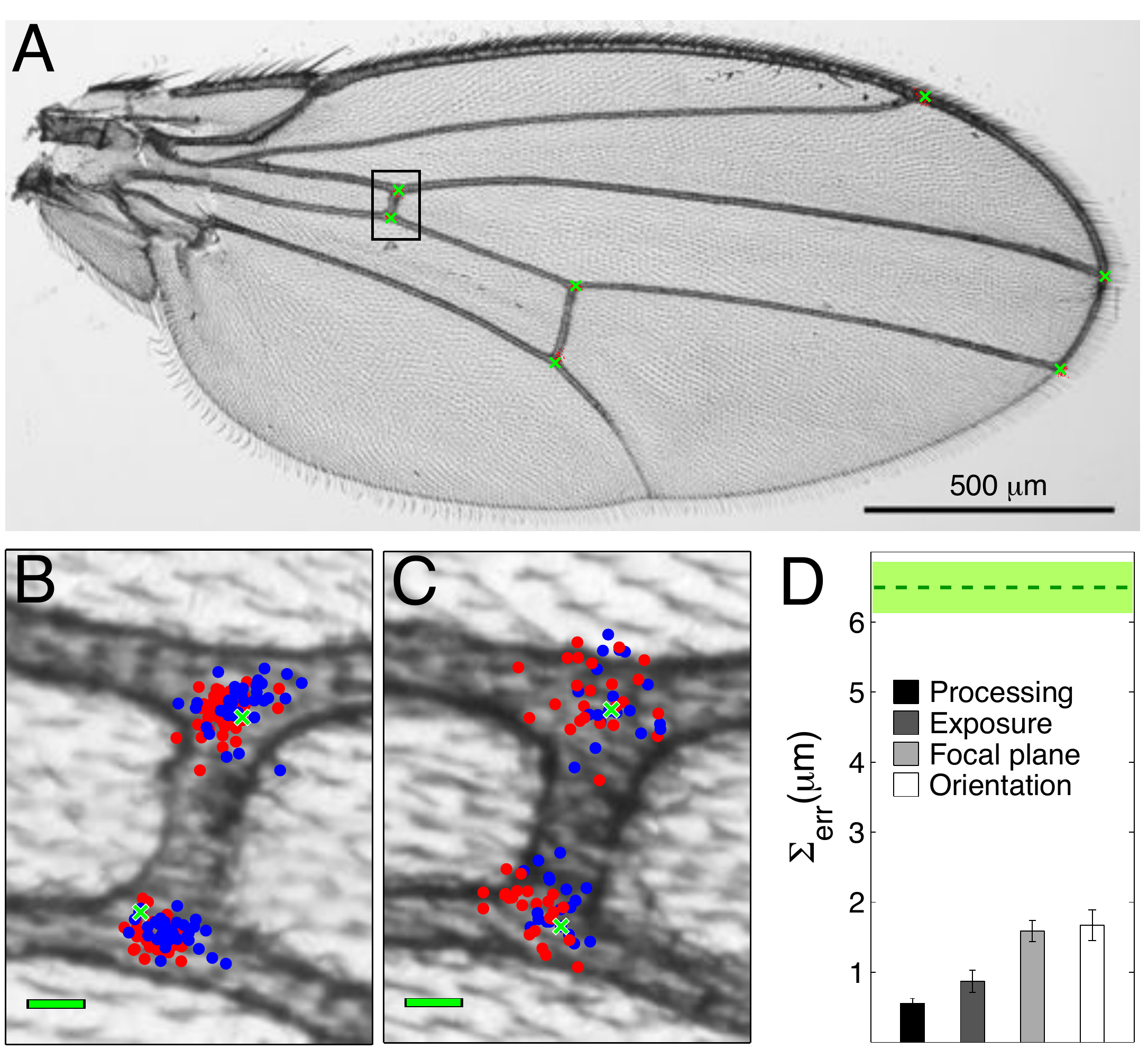}
\caption{{\bf Wing vein pattern measurements. (A)} Right wing of a male {\it Drosophila} adult fly. The coordinates of seven wing vein crossings (landmarks) quantify principal pattern features. Green crosses indicate landmark positions of the depicted wing, respectively in all panels. Close-up of the two most proximal landmarks (black rectangle) with the procrustes superimposed landmark distribution from right wings of {\bf (B)} 41 flies raised under optimal conditions (OreR raised at 18C) and {\bf(C)} 22 flies from a natural population (caught in Cartagena, Colombia, raised at 22C). Individual landmark locations are shown for females and males in red and blue, respectively. Green scale bar represents linear dimension of an average wing cell. The distributions show the smallest (in B) and largest (in C) landmark variation observed. {\bf(D)} Various systematic measurement errors $\Sigma_{\rm err}$ on the location of landmark locations~\cite{SM}. Horizontal green line is the average spread of landmark distributions ($\braket{\Sigma}=6.5\pm0.7\mu m$) across male wings of all fly lines (N=143), see text. }
\label{fig1}
\end{figure}

Wing morphogenesis is controlled by a multitude of genes~\cite{Carreira:2011} as well as by environmental factors such as developmental temperature~\cite{Debat:2003}. Here we quantify how the final wing architecture is modified across the naturally occurring genetic variation in wild-type fly populations as well as by developmental temperatures within the viable range. Our measurements show that, independent of the particular genetic makeup and the temperature set-point, the final wing pattern precision is at the level of half the linear dimension of a single wing cell. For some specific sets of genetic and temperature conditions, the system performs as reproducibly as it is precise; for no set is the reproducibility worse than the linear cell size. Given that a single cell is the minimal physical unit at which tissue patterning can be realized, our findings suggest that wing patterning is optimized to operate at the physical limit of the system.

Naturally, the two most prominent features characterizing a wing are its size and its shape or pattern. Notably, the wing pattern is scalable and independent of size variation~\cite{Breuker:2006} which allows us to quantify only variations in wing vein patterning; size variations depend on nutritional intake which is difficult to control~\cite{footnote_wing_size}. Size invariance allows us to apply a succession of affine transformations on each wing from a given data set, known as Procrustes transformations~\cite{Bookstein:1991}, in order to only quantify variations in the principle pattern features~\cite{Klingenberg:2000, Zimmerman:2000} (see Supplementary Material~\cite{SM}). Pattern features are captured by a configuration of seven landmark coordinates $\{x_i,y_i\}$, which demarcate wing vein crossing points (Fig. 1A); wing size is measured by the Euclidean norm $S$ of the configuration (Fig. S1~\cite{SM}).  Deviations from a mean reference pattern are minimized in an iterative procedure.  The average linear (1d) variation in the spatial landmark location is captured by the invariant trace of the covariance matrix $C$ of the seven landmark coordinates (i.e. $7\times2$ dimensions):
$$\textstyle \Sigma = \braket{S}\sqrt{\frac{1}{14}\Tr(C)}.$$ 
Importantly, this transformation reconstructs a mean wing of size $\braket{S}$, which allows us to measure departures from the mean pattern in absolute units and compare these to a physically relevant length scale such as the size of a wing cell~\cite{footnote_cell_size}. Additionally, there are systematic differences in the vein patterning between sexes~\cite{Carreira:2011}, and thus we assess pattern variations in males and females separately. Particularly surprising is the accuracy with which transformed right wings overlap in the least (Fig. 1B) and most (Fig. 1C) variable landmark distributions among all examined fly lines. In both cases the spatial extent of the landmark distributions is comparable to the average linear dimension of an individual wing cell (i.e. $\sim\!13\mu m$ \cite{SM}). Moreover our experimental error in determining the location of individual landmarks only represents a small fraction if that~\cite{SM}, i.e. $\Sigma_{\rm err} = 1.67\pm0.22\,\mu m$ (Fig. 1D), implying that we are measuring mostly true biological variation. Hence, the spatial variation of the landmarks demonstrates a remarkable level of scale invariance and a highly conserved vein pattern.

To measure {\it precision} and {\it reproducibility} requires us to superimpose pairs of transformed wings and to quantify their spatial variations in terms of the differences between landmark locations. Precision is assessed from the differences in landmark locations $(\Delta_{\rm L-R})$ between the left and right wing in the same fly, while reproducibly is computed by the differences in landmark locations $(\Delta_{\rm L-R'})$ of a pair of randomly chosen left and right wing from the entire fly population (note the prime marking different individuals). To obtain the average landmark within-individual variance (${\Sigma_{\Delta_{\rm L-R}}}$), we calculate the covariance matrix $C$ of the differences in the landmark coordinates of pairs of wings  $\left(\{x_i^{\rm L}-x_i^{\rm R}, y_i^{\rm L} - y_i^{\rm R}\}\right)$.  Therefore $\Sigma_{\Delta}$ corresponds to the average variation in the distance between landmarks and is statistically related to the 1-d variation of landmark locations defined above (Fig. S1~\cite{SM}). Analogously, average landmark individual-to-individual variations (${\Sigma_{\Delta_{\rm L-R'}}}$) are computed from random pairs of wings in the population. Note that high precision or reproducibility are measured as low variations $\Sigma$.

\begin{figure}[t!]
\centering
\includegraphics[width = \linewidth]{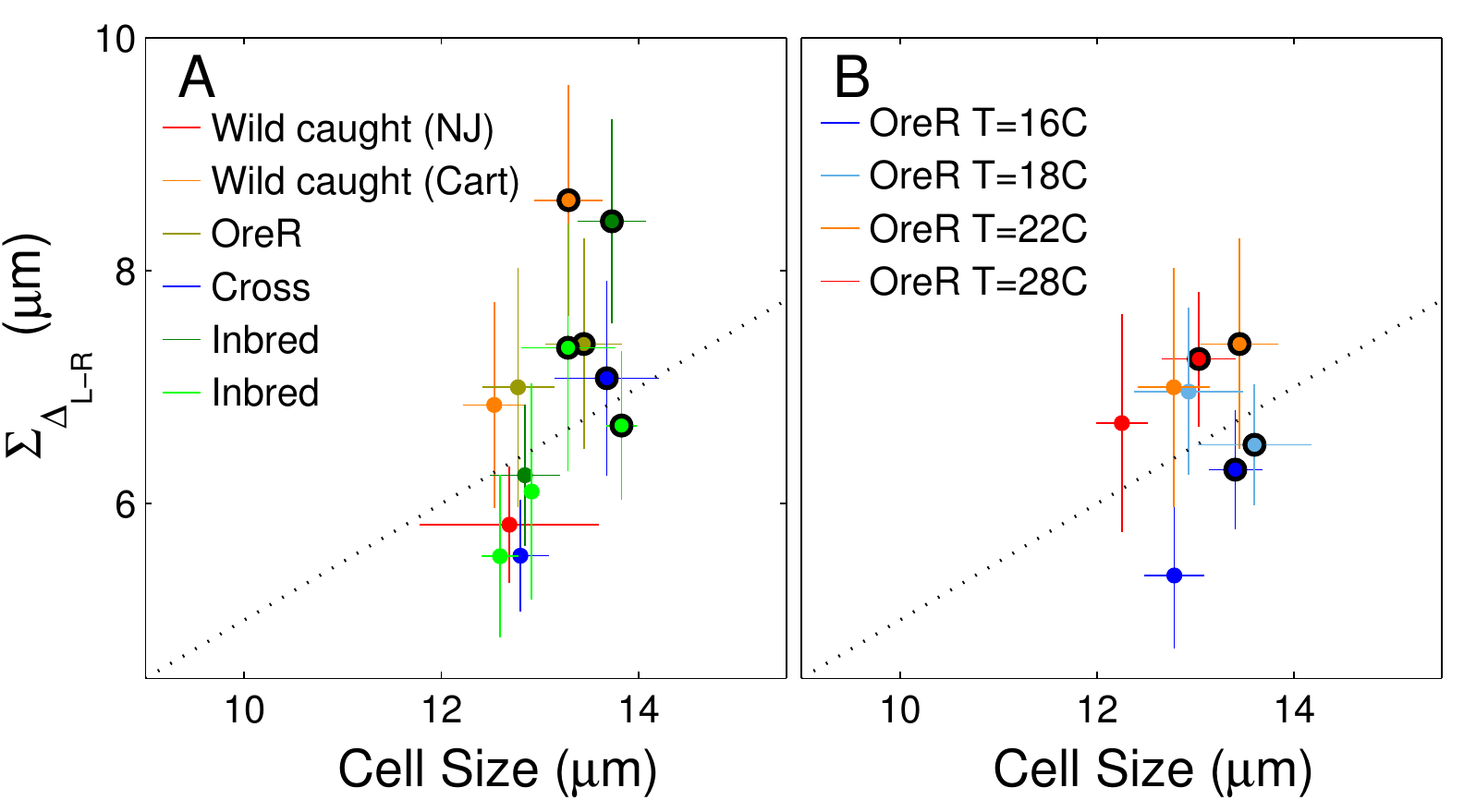}
\caption{\textbf{Spatial precision of vein patterns under genetic and environmental variation.} 
For fly populations of different genetic makeup (A) and at different growth temperatures (B) the spatial within-individual landmark variation ($\Sigma_{\Delta_{\rm L-R}}$), a measure of precision, is shown as a function of the average linear wing cell size of the population. Females are indicated by black circles. In (A) all flies are raised in a controlled environment at 22$^\circ$C; in (B) identical genetic background (OreR) is used. In both cases precision of the vein pattern is unaffected and remains at half a cell size (dotted lines have slope 0.5).}
\label{Fig3}
\end{figure}

We first assess the fidelity of the patterning process by quantifying its precision in different genetic (Fig. 2A) and temperature (Fig. 2B) scenarios from the naturally available ranges.  In particular, we utilize populations of flies with different levels of genetic heterogeneity: i.e. homogenous populations of individuals with identical genotype (called inbreds), intermediate heterogeneity (crosses of inbreds or laboratory-bred Ore-R flies) and wild-caught populations. All fly populations are raised in identical environmental conditions and at room temperature (i.e. $22^\circ$C), except when effects of developmental temperature are directly explored, for which we chose population-specific temperature set-points within the viable range  (i.e. $14^\circ$C to $30^\circ$C). A priori, one expects that changes in genetic makeup lead to changes in precision \cite{Debat:2003}, and that non-optimal environmental conditions such as extreme temperatures are stressful to the patterning process and reduce its precision \cite{Breuker:2006}. Surprisingly, the precisions for each of our fly lines cluster around half a cell size (denoted as dotted lines in Fig. 2), independent of genetic heterogeneity or highly non-optimal growth temperatures. For the landmark distributions in Fig. 1 this result indicates that the system-intrinsic error in positioning landmarks is of the order of half a cell \cite{footnote_cell_size}. The remaining spread of the landmark distribution around the mean location must therefore reflect systematic effects which we will now interpret through measurements of pattern reproducibility. 

%Given that the vein pattern is controlled by multiple genes, one expects that increasing the genetic differences in a fly population reduces pattern reproducibility in the ensemble.  We see this effect in our fly populations, but surprisingly, the bounds on reproducibility are between the linear dimension of a wing cell and half of that (Fig. 2A). In addition, the precisions for each of our fly lines also cluster around half a cell size, independent of genetic heterogeneity (Fig. 2B). A similar behavior is observed when patterning proceeds in non-optimal environmental conditions. A priori, one expects that extreme temperatures within the viable range are stressful for the patterning process and reduce both precision and reproducibility. Indeed, in fly populations raised at different temperature set points we observe a temperature optimum at $18^\circ$C, above and below of which reproducibility is reduced (Fig. S4). However, the lower and upper bounds on reproducibility are again at the level of a single wing cell and at half of that, respectively, and the precision of the process remains at half a cell size. Together these results indicate that even under strong genetic heterogeneity or highly stressful growth temperatures the performance of the patterning process is only moderately altered; the precision is unaffected by these conditions, and reproducibility decreases at most twofold. 

The simplest way to assess pattern reproducibility is to directly contrast it with the precision measurements in a scatter plot for both genetic and temperature scenarios  (Fig. 3). The observed constancy of patterning precision suggests that spatial decisions are matched to the linear dimension of a wing cell. Therefore, to compare different fly lines, the measurement are normalized by the average cell size of the respective fly line (see Fig. S2 \cite{SM}). Thus variations due to differences in the inherent length scale are excluded and, as a result, the spread on the precision axis tightens (converging to half a cell size), but not on the reproducibility axis.

In each case, we observe conditions for which the wings in different animals are as similar to each other as the wings within an individual. These conditions are identified by data that cluster along the diagonal $\Sigma_{\Delta_{\rm L-R'}}=\Sigma_{\Delta_{\rm L-R}}$, implying that the same high fidelity patterning program must operate in every single wing, be it within or across individuals. In the genetic case (Fig. 3A), these ideal conditions are attainable only for populations comprising individuals with identical genetic composition (i.e. inbreds, raised in the same environment). For these flies, the vein pattern cannot identify whether two random wings in a population stem from the same individual or not, indicating that wing patterning proceeds independently in left and right wings of a given fly. 
   
\begin{figure}[t!]
\centering
\includegraphics[width=\columnwidth]{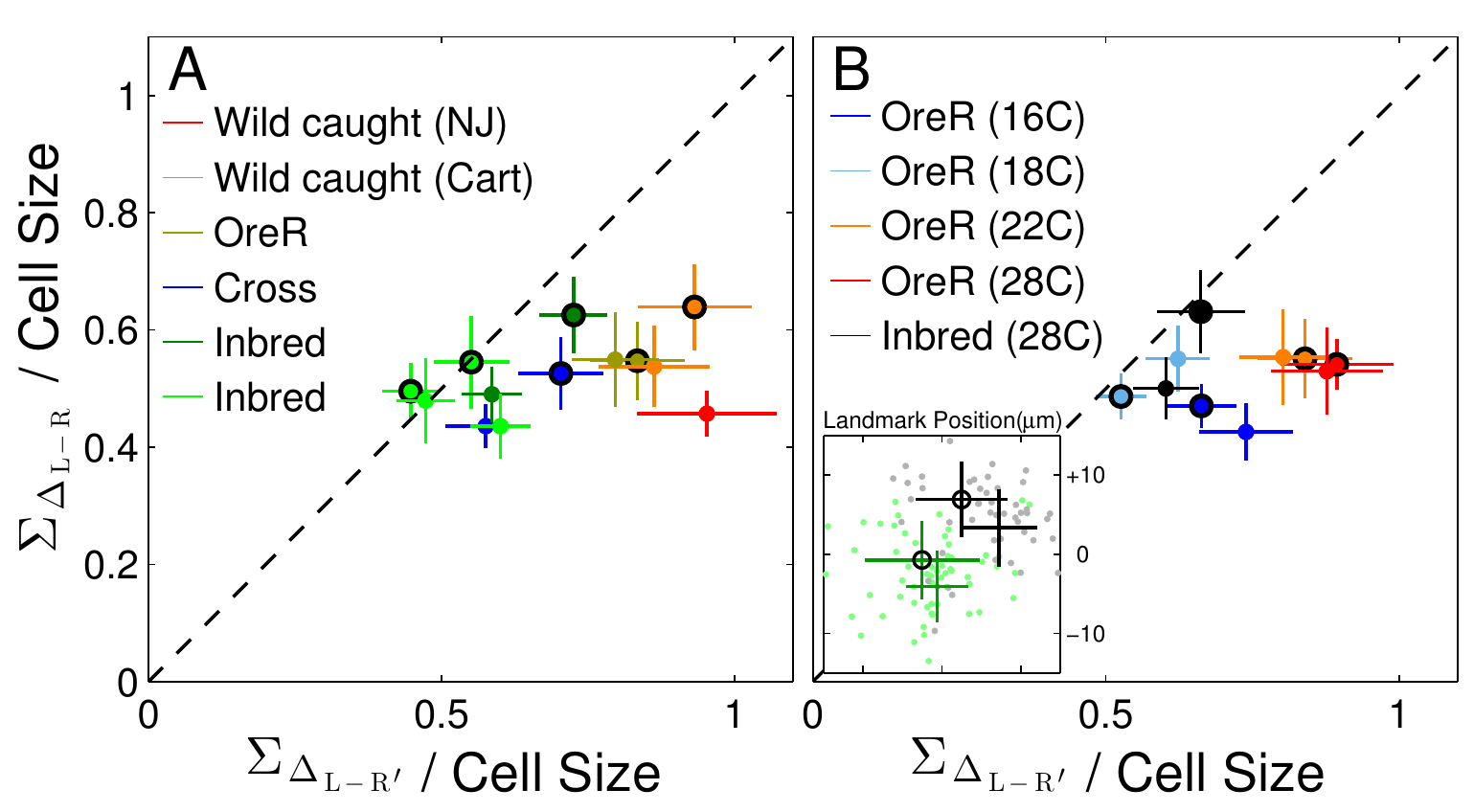}
\caption{{\bf Precision versus reproducibility under varying genetic (A) and temperature (B) conditions.} Precision (y-axis) and reproducibility (x-axis) are measured in units of the relevant physical unit, the linear wing cell size. As expected, flies are on average more precise (symmetric) than reproducible (dashed line for identity). However, under controlled genetic and environmental conditions (in the most inbred lines in (A) and at 18$^\circ$C in (B)) wings are as reproducible as they are precise, i.e. the left and right wing of the same fly are as similar to each other, as they are to the wings of another fly in a given fly line. Thus, pattern reproducibility equals pattern precision. Inbreds raised at 28$^\circ$C (black data in (B)) show no reduction in reproducibility. Inset in (B) shows the net difference in the location of landmark 2 between two inbred populations raised at 22$^\circ$C (light green data in (A)) and at 28$^\circ$C (black data in (B)).}
\label{fig1}
\end{figure}

Since individuals from a homogeneous population have nearly indistinguishable vein patterns, the decrease in reproducibility in a heterogeneous fly population (spread on x-axis in Fig. 3A) must result from an increasing number of genetic compositions, which modify the mean landmark configuration and hence widen the distribution of landmark locations. Each landmark configuration corresponds to a particular genetic composition, which when observed in an inbred population, would generate a pattern variation that is at the above observed precision optimum. However, when multiple landmark configurations are mixed in a population, the reduced reproducibility results from fluctuations caused by changes in the means of the different configurations, thus increasing the overall spread of the population distribution. Remarkably, the bounds on reproducibility are surprisingly small; they are as large as only a single wing cell and as low as half of that, rendering all vein patterns identical at that limiting resolution.  

In the temperature case (Fig. 3B), we observe an optimum at $18^\circ$C for which vein patterns are as precise as reproducible even in a fly line that contains genetic heterogeneity. Reproducibility decreases for lower and for higher temperatures, possibly due to temperature-induced stress that affects the fidelity of the patterning process. However, contrary to that intuition, the constant precision across all temperature set-points indicates that the large spread along the reproducibility axis more likely results from temperature induced amplification of the genetic differences in that fly population rather than from a decline in patterning fidelity. We can test this conjecture directly in an inbred fly population where two subsets are raised at two non-optimal temperature conditions (e.g. $22^\circ$C and $28^\circ$C). In these populations temperature has no effect on the overall variability in landmark positions: wings in the same animal are as similar as those in different animals (green and black data in Fig. 3A and 3B, respectively). For these conditions, the vein pattern is indeed generated with half a cell precision, but two distinct mean landmark configurations are observed (Fig. 3B, inset, and Fig. S3 \cite{SM}). Each configuration is generated with half a cell precision, and the overall reproducibility remains within a single cell. 

%%%

Together, our observations point to a patterning program which operates analogously to a high fidelity production line, independently in each wing. The performance has a precision of half a cell, and it is stable under the range of naturally occurring genetic variations and environmental temperatures. Differences in either condition can only lead to systematic shifts of the means of the landmark locations, and the spatial limit of these shifts is at the level of the size of a single wing cell, rendering landmark configurations remarkably conserved. A genetically heterogeneous population is thus akin to an ensemble of production lines, each of which runs in identical ambient conditions and differs by its genetic composition. In all explored sets of genetic and environmental conditions, the key factor inducing changes in the vein pattern is an increasing genetic availability of landmark configurations in the fly population: the higher the number of configurations, the lesser the observed pattern reproducibility.

From a functional point of view the single-cell invariance of the vein pattern in both genetic and temperature scenarios might be surprising. However, symmetry between the left and right wing of an individual is functionally important for proper steering during flight~\cite{Lehmann:2009} and for successful courtship~\cite{Pavkovic:2011}. Hence, some evolutionary pressure is expected on precision. Thus, while the functional role for single-cell reproducibility of the vein pattern is unknown, it is likely related to the animal's overall flight control. The vein pattern is geometrically important for aerodynamic properties~\cite{Combes:2003}, while the ratio of wing size to body size relates to overall flight capability~\cite{David:1994}. Therefore single cell reproducibility might reflect the necessary level of scale invariance to ensure perfect aerodynamic control and the necessary level of size matching to maintain flight ability despite differences in body sizes. 

We have primarily focussed on pattern variability and neglected considerations on the variability of wing size, mainly due to technical difficulties as mentioned above.  Our results on patterning reproducibility suggest that wings form independently and no left--right communication is necessary (see also Fig. S3~\cite{SM}). On the other hand it has been shown that hormonal signaling coordinates size determination of fly organs and, in particular, achieves similar wing size in the same animal~\cite{Garelli:2012}. But it remains possible that such a strategy solely ensures proper matching of organ proportions and is not required for final wing size determination and maintenance of bilateral symmetry~\cite{Lawrence:2000}.

Overall, our results support a straightforward strategy for the maintenance of bilateral symmetry during the wing generation process. Wing morphogenesis begins with as few as 6 cells in a 3h old embryo~\cite{Wieschaus:1976} that undergo 12-13 divisions within 6 days to generate a disc in the fly larva with $\sim\!50,000$ cells (Table I).  At this point cells no longer divide, and cell--cell connectivity properties~\cite{Classen:2005} and global mechanical deformations~\cite{Aigouy:2010} shape the adult wing. As long as the seed cells on the left and the right sides of the developing wing structures are symmetrical, patterning can proceed independently with high fidelity on both halves of the fly and result in highly symmetric wings. This suggests that all spatial decisions in the previous layers of the wing formation process should be performed with a spatial precision better than that of the final product, i.e. a single cell.   Indeed, a similar situation has been observed in the early embryo, where the classical example of reproducibility is the location of the first morphologic mark on an otherwise uniform sheet of cells, i.e. the cephalic furrow~\cite{Namba:1997}, whose location is also reproducible at the level of half a cell size~\cite{Dubuis:2013a, Liu:2013}. This connection indicates that the spatial reproducibility of morphogenesis in the fly may be maintained throughout the entire 10 days of development.  

In principle, morphogenetic processes in the fly could have reproducible outcomes by measuring at each stage the size of the relevant local ``building block" of the pattern (i.e. an individual cell) and determining position with spatial precision of half of that unit's size (Table I). This suggests that an error of half the size of the building block is a sufficient strategy for generating and maintaining spatial reproducibility from one patterning layer to the next. In particular, it seems to be sufficient to reproducibly generate a complex pattern comprised of as many as $\sim\!20000$ such units in the {\it Drosophila} wing. It is important to test how these constraints affect current models for growth and pattering at different stages of the wing formation process~\cite{Kauffman:1978, Shraiman:2005, Hufnagel:2007}.

%%%
\begin{table}[t!]
\caption{\label{tab:table1}Size comparison of features in wings and embryos.}
\begin{ruledtabular}
\begin{tabular}{ld}
Wing features (developmental time) & \mbox{Cells}\\
\hline
Wing disc, embryo (3h)~\cite{Wieschaus:1976} & \sim\!6 \\
Wing disc, embryo (10h)~\cite{Bate:1991}& \sim\!24 \\
Wing disc, larva (24h)~\cite{Garcia-Bellido:1971} & \sim\!50 \\
Wing disc, pupa (6d)~\cite{Garcia-Bellido:1971} & \sim\!50000\\
Wing blade, adult (10d) & \sim\!20000\\
\hline
Features (wing 10d; embryo 3h)&\mbox{Size ($\mu$m)}\\
\hline
Wing length&{\sim\!1900}\\
Wing cell size&13.0\pm0.7 \\
Wing landmark reproducibility&6-12.5 \\
Wing landmark precision&6.5\pm1.3 \\
Embryo length &{\sim\!490}\\
Embryo cell size~\cite{Dubuis:2013a}&8.2\pm1.0 \\
%C\footnote{Some tables require footnotes.}
Embryo spatial reproducibility~\cite{Dubuis:2013a}&4.2\pm0.7 \\
\end{tabular}
\end{ruledtabular}
%\footnotetext[1]{ See Ref. \cite{Dubuis:2013a}, }.\footnotetext[2]{ See Ref. \cite{Wieschaus:1976}}. \footnotetext[3]{ See Ref. \cite{Bate:1991}}. \footnotetext[4]{ See Ref. \cite{Garcia-Bellido:1971}}.
\end{table}

Could optimized reproducibility be a general feature of morphogenesis? Our findings suggest that it might at least be the case for the maintenance of symmetric features during developmental growth. Interpreting variations in absolute units allows us to recognize that wing patterning runs with a precision of less than a single cell. Because it is individual cells that make fate determining decisions, a cell is arguably the minimal physical unit at which tissue patterning  can be realized. We identify here a signature of optimization in developmental processes, which seemingly perform at their physical limit in generating patterns. Our work thus represents a necessary first step toward an understanding of reproducibility at non-equilibrium.

We thank P. Andolfatto, C. Broedersz, A. Kumar, A. Sgro, and M. Tikhonov for technical help and comments. This work was supported by National Institutes of Health Grant P50 GM071508 and by Searle Scholar Award 10-SSP-274 to Thomas Gregor.

\end{document}